\documentclass[aps,prl,twocolumn,amsmath,amssymb, superscriptaddress,floatfix]{revtex4-2}

\usepackage[utf8]{inputenc}
\usepackage{xcolor}
\usepackage{pst-node}
\usepackage{graphicx}
\usepackage{amsmath, amssymb, amsthm}
\usepackage{mathtools}
\usepackage{hyperref}
\usepackage[shortlabels]{enumitem}
\usepackage{dsfont}
\usepackage{amsthm}
\usepackage{amssymb}
\usepackage{amstext}
\usepackage{amsfonts}
\usepackage{nicefrac}
\usepackage{extarrows} 
\usepackage{graphicx}
\usepackage{relsize}

\newcommand{\ket}[1]{\vert #1 \rangle}
\newcommand{\bra}[1]{\langle #1 \vert}
\newcommand{\mat}[1]{\underline{\underline{#1}}}

\newcommand{\calB}{{\mathcal B}}

\begin{document}

\title{
Block entropy area based
non-local fermionic mode optimization\\ with gradient disentanglers}

\author{Mikl\'os Antal Werner}
\email{werner.miklos@wigner.hu}
\affiliation{%
Strongly Correlated Systems Lend\"ulet Research Group,
Wigner Research Centre for Physics, H-1525, Budapest, Hungary
}%

\author{Gero Friesecke}
\email{gf@ma.tum.de}
\affiliation{%
  Department of Mathematics, Technical University of Munich, Germany
}%

\author{Andor Menczer}
\affiliation{%
Strongly Correlated Systems Lend\"ulet Research Group,
Wigner Research Centre for Physics, H-1525, Budapest, Hungary
}%

\author{Korn\'el Kap\'as}
\affiliation{%
Strongly Correlated Systems Lend\"ulet Research Group,
Wigner Research Centre for Physics, H-1525, Budapest, Hungary
}%

\author{{\"O}rs Legeza}
\email{legeza.ors@wigner.hu}
\affiliation{%
Strongly Correlated Systems Lend\"ulet Research Group,
Wigner Research Centre for Physics, H-1525, Budapest, Hungary
}%
\affiliation{%
Parmenides Stiftung, Hindenburgstr. 15, 82343, P{\"o}cking, Germany
}
\affiliation{%
Dynaflex LTD, Zr{\'i}nyi u 7, 1028 Budapest, Hungary
}

\date{\today}

\begin{abstract}
We introduce a systematic block entropy area based mode optimization algorithm for many-body quantum states of interacting fermions represented by matrix product states. From the gradient of
a global cost function, the block entropy area, a long-ranged, non-interacting effective disentangler Hamiltonian is formed. We then simulate the time-dependent Schrödinger equation driven by the disentangler Hamiltonian by employing the time-dependent variational principle based on projector splitting, and minimize the cost function. The combination of the density matrix renormalization group with this gradient-based entanglement minimization forms an efficient low-rank iterative ground-state algorithm that also provides an optimized single-particle basis for matrix product state representation. We demonstrate the method on two-dimensional lattice models of interacting fermions and the Fe${_4}$S${_4}$ cluster, and show its robustness and superiority over earlier protocols using nearest-neighbor mode rotations and reorderings.
\end{abstract}

\maketitle
\emph{Introduction\,– } The solution of the time-independent Schrödinger equation for quantum systems built from many constituents, i.e., finding eigenstates of the many-body Hamiltonian, has challenged researchers since the earliest days of quantum mechanics~\cite{Fetter-2012, Negele-2018}. The dimension of the Hilbert space of many-body quantum states scales exponentially with the number of constituents, thereby restricting numerically exact diagonalization to rather small models and requiring approximations for most systems of interest~\cite{Schollwock-2005, Foulkes-2001, Held-2007, Georges-1996, Bartlett-2007, Metzner-2012, Carleo-2017}.  For quantum states with moderate entanglement, tensor network state (TNS) methods provide an efficient and controlled approximation~\cite{White-1992,Schollwock-2011,Orus-2014,Szalay-2015,Cirac-2021,Verstraete-2023}; here, the required number of wave-function parameters scales exponentially with the amount of entanglement rather than with the number of microscopic constituents~\cite{Schuch-2008,Eisert-2010,Stoudenmire-2012}. However, for states with larger entanglement, e.g., for critical models or Hamiltonians with long-ranged couplings, the high computational requirements can prohibit calculations at the desired accuracy threshold, even if the most advanced hybrid CPU/GPU high-performance computing (HPC) solutions are employed~\cite{Menczer-2024b, Menczer-2024c, Menczer-2024a, Xiang-2024}. Decreasing the entanglement of the target state by an optimal choice of the computational basis can therefore tremendously boost simulations.
\begin{figure}[!h]
    \includegraphics[width=0.42\textwidth]{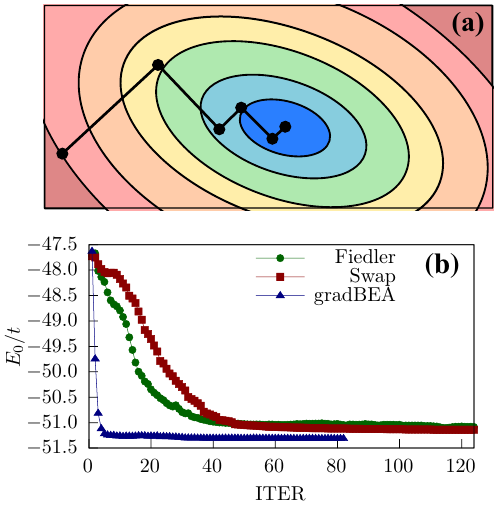}
    \caption{\textit(a) Illustration of the steepest descent algorithm. After setting the search direction to the gradient of the cost function (shown as a contour plot) at a parameter point, the point is updated by minimizing the cost function along the straight search line. \textit{(b)} Ground state energy of the $10 \times 10$ sized spinless tt'V model at $t'/t = 0.4$ and $V/t = 0.8$ as a function of mode optimization iterations at bond dimension $D=80$. The method introduced in our work (blue triangles) shows an order of magnitude faster convergence compared to the local methods using Fiedler vector-~\cite{Krumnow-2016} or systematic swap-gate-based~\cite{Friesecke-2024} reorderings. The best (lowest) ground state energy in our work $E_0/t = -51.313$ is also significantly below the swap-gate and Fiedler vector-based methods ($E^\mathrm{swap}_0/t = -51.146$,  $E^\mathrm{Fiedler}_0/t = -51.104$). \label{fig:Descent_and_Spinless}}
\end{figure}

The role of the single-particle basis in the efficiency of the density matrix renormalization group (DMRG) algorithm, a subclass of TNS methods, has been studied from various aspects in the past decades ~\cite{Xiang-1996,Rissler-2006,Legeza-2006b,Murg-2010a,Olivares-2015,Keller-2014,Sharma-2014}.
Along this path,
entanglement-based fermionic mode optimization was first introduced in Ref.~\cite{Krumnow-2016}, where two-mode local (nearest-neighbor) optimization microsteps 
via the bipartition R\'enyi entropy~\cite{Vidal-2003,Legeza-2003b,Verstraete-2006} at a given bond, together with a
heuristic global reordering of modes based on the 2-mode mutual information matrix were employed based on the so-called Fiedler vector~\cite{Barcza-2011}. While the method shows robust initial convergence, undamped and uncontrolled oscillations often appear close to the observed optimum~\cite{Krumnow-2021, Friesecke-2024}. 
Nevertheless, the numerical power of the algorithm has been successfully demonstrated on various systems like lattice models~\cite{Krumnow-2016,Krumnow-2019,Krumnow-2021,Menczer-2024a}, molecules~\cite{Petrov-2023,Mate-2023}, and nanotubes~\cite{Moca-2020}, introducing the concept of Rényi entropy optimized molecular orbitals (REMO)~\cite{Petrov-2023}.

As a significant improvement, in Ref.~\cite{Friesecke-2024} a global version of this idea was introduced: a single, entanglement-based global cost function, the block entropy area (BEA) is optimized over the whole manifold of fermionic modes.
Traversing through the whole phase space of two-mode rotations is guaranteed by systematic rigorous reorderings formed by layers of swap gates.
However, the scrambling caused by
global swap-gate re-orderings often lead to undamped oscillations~\cite{Friesecke-2024,Werner-2026}. Those can be avoided by simply rejecting bad reorderings~\cite{Li-2025}; however, that increases the numerical cost significantly and can also reduce the convergence rate.

In this work, we keep the idea of optimizing the BEA, introduced in \cite{Friesecke-2024}, but propose a mode optimization protocol that is fundamentally different from former approaches discussed above. Instead of restricting optimization to nearest-neighbor pairs, we calculate the gradient of the BEA,
introduced in Ref.~\cite{Friesecke-2024},
which involves all generators of the rotation group, and use the steepest descent method for minimization~\cite{Nocedal-2006}, which is a standard and powerful tool in numerical analysis (see Fig.~\ref{fig:Descent_and_Spinless}a). From the gradient, a non-interacting disentangler Hamiltonian is formed that drives an effective time evolution to reduce entanglement. Here time parametrizes the one-parameter rotation subgroup generated by the gradient. The new method, dubbed gradBEA, is free of reorderings and is genuinely \emph{non-local}, as the disentangler Hamiltonian contains couplings between every mode pair. 
Our numerical results on two-dimensional lattice models and the Fe$_4$S$_4$ cluster show that the method leads to faster convergence and results in significantly lower ground state energies than the above-discussed protocols, while the numerical requirements (runtime per iteration) are only slightly increased.

\emph{Algorythmic details\,– } The Hamiltonian of a fermionic system with two-body interactions has the general form 
\begin{equation}
    \hat{H} = \sum_{ij} \sum_{s} t^{(s)}_{ij} c^\dag_{i s} c_{j s} + \sum_{ijkl} \sum_{s,s'} v_{ijkl}^{(ss')} c^\dag_{i s} c^\dag_{j s'} c_{k s'} c_{l s} \; ,  \label{eq:Hamiltonian_spinful}
\end{equation}
where $c^\dag_{is}$ ($c_{is}$) creates (annihilates) a fermion at mode position $i \in [1, 2, \dots N]$ and spin $s \in  \lbrace \uparrow, \downarrow \rbrace$ for spinful models~\footnote{We note that the form of the Hamiltonian has to be generalized for higher spin ($S>1/2$) models.}. The expression \eqref{eq:Hamiltonian_spinful} is, however, not unique: A unitarily equivalent family of Hamiltonians having the same general form can be written using the transformed fermion operators $\tilde{c}^\dag_{j s} = \sum_{i} U^{(s)}_{ij} c^\dag_{i s}$ and the accordingly rotated matrix elements $\tilde{t}^{(s)}$ and $\tilde{v}^{(ss')}$. Here $U^{(s)}_{ij}$ denotes the two $N\times N$ unitary matrices that transform the operators of spins $\uparrow$ and $\downarrow$.  The equivalent Hamiltonians are not equally suitable if approximations are used. The matrix product state (MPS) wave function, optimized within DMRG~\cite{White-1992, Schollwock-2005, Schollwock-2011, Szalay-2015}, is the most powerful in a basis where the entanglement between fermionic modes is small~\cite{Schuch-2008,Eisert-2010,Stoudenmire-2012}. Suppression of entanglement in parallel with the ground state energy is the main goal of our protocol. We note that in the context of quantum chemistry, for example in the self-consistent field (SCF) method~\cite{Roos-1980}, mode optimization can also leave the active space spanned by the original $N$ fermionic modes~\cite{Zgid-2008c,Legeza-2025}. However, similarly to former entanglement-based mode optimization techniques, we consider only rotations within the $N$-dimensional single-particle space spanned by the original modes.  

MPS wave functions form a special class of tensor network states with linear (chain) tensor topology~\cite{Ostlund-1995,Schollwock-2011},
\begin{equation}
  \ket{\Psi^{\mathrm{MPS}}} = \sum_{\lbrace\sigma_1 \sigma_2 \dots \sigma_N \rbrace} \mat{M}^{[1] \sigma_1} \mat{M}^{[2] \sigma_2} \dots \mat{M}^{[N] \sigma_N} \ket{\sigma_1 \sigma_2 \dots \sigma_N} \; , \label{eq:MPS}
\end{equation}
where $\sigma_i \in \lbrace 0,\,\downarrow,\,\uparrow,\,\uparrow\downarrow \rbrace$ stand for local state at mode position $i$. The computational basis $\ket{\sigma_1 \sigma_2 \dots \sigma_N}$ is formed as the tensor product of 4-dimensional local Hilbert spaces, while the anticommutation relations of fermion operators are ensured by the Jordan-Wigner transformation~\cite{Jordan-1928, Solyom-2007Book1}. The maximal allowed size of matrices $\mat{M}^{[i] \sigma_i}$ is the so-called bond dimension $D$ that controls the expressiveness of the ansatz. DMRG approximates the ground state $\ket{\Psi_0}$ of $\hat{H}$ in an iterative variational protocol using the MPS ansatz. The error caused by bond-dimension truncation between sites $(i,i+1)$ is strongly related to the Schmidt decomposition~\cite{Schmidt-1907, Schollwock-2005} of the state at the partitioning of sites as $A: [1,2,\dots,i]$ and $B: [i+1,\dots,N]$,
\begin{equation}
\ket{\Psi_0} = \sum_{\alpha}\lambda_{\alpha}\ket{\alpha}_A \ket{\alpha}_B \, , \label{eq:schmidt_decomp}
\end{equation}
where $\lbrace \ket{\alpha}_A, \ket{\alpha}_B \rbrace$ are the pairs of Schmidt states in the subsystems $A$ and $B$, and the amplitudes $\lambda_\alpha\ge0$ are the Schmidt-values for which $\sum_\alpha \lambda_\alpha^2 = 1$ by normalization. Keeping only the $D$ largest Schmidt values provides an optimal approximation of the state whose error is measured by the truncation error $\epsilon_{\mathrm{Tr}} = \sum_{\alpha>D} \lambda_\alpha^2$. 

\begin{figure}
    \includegraphics[width=0.48\textwidth]{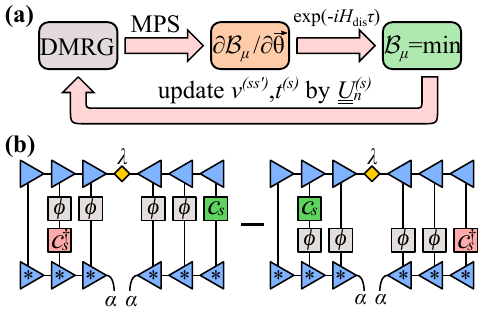}
    \caption{\textit{(a)} Iteration loop of the gradient-based mode optimization algorithm. From the MPS wavefunction produced by DMRG, we determine the gradient of $\calB_\mu$, and construct the disentangler Hamiltonian $\hat{H}_\mathrm{dis}$. Time evolution with this decreases $\calB_\mu$, and is minimized at some time $\tau^*$, is used to construct $\mat{U}^{(s)}_n$ in Eq.~\eqref{eq:U_n}. \textit{(b)} Tensor diagram of $\Xi^{(s,kl)}_{\alpha \alpha}$ used in Eq.~\eqref{eq:palpha_derivative} for $\lbrace k=2,l=6 \rbrace$. The yellow diamond is the diagonal matrix containing the Schmidt values $\lambda_{\alpha}$, the blue triangles are in the left and right are left- and right-canonical MPS matrices~\cite{Schollwock-2011}, and squares stand for single-site operators, where $\phi=(-1)^{n_\uparrow + n_\downarrow}$ is the Jordan-Wigner sign.\label{fig:GradBEA_graph}}
\end{figure}

Transformation of modes by a unitary $U^{(s)}_{ij}$ also transforms the Schmidt spectrum $\lambda_\alpha$ and the truncation error. Direct minimization of the sum of truncation errors -- a naive cost function candidate --  over bonds is numerically ill-posed due to the hard cutoff in the definition of $\epsilon_{\mathrm{Tr}}$. Therefore, we use the R\'enyi entropies $S_\mu^{(i)} = \frac{1}{1-\mu} \ln \left(\sum_\alpha \lambda_\alpha^{2\mu} \right)$ to quantify entanglement~\cite{Verstraete-2006}, where $0<\mu<\infty$ parametrizes a family of entropy measures (with $\mu \rightarrow 1$ being the usual von Neumann entropy). The superscript $(i)$ indicates the rightmost site of the left block $A$. The sum of the entropies over bonds defines the cost function \cite{Friesecke-2024, Werner-2026} 
\begin{equation}
    B_\mu = \sum_{i=1}^{N-1} S^{(i)}_{\mu} \; , \label{eq:BEA_definition}
\end{equation}
that we aim to minimize. The optimal mode set depends on the value of $\mu$; in this work we use $\mu=0.5$, similarly to Refs.~\cite{Krumnow-2016,Friesecke-2024,Werner-2026, Li-2025}. Instead of the hopelessly difficult direct minimization over unitaries $U^{(s)}$, we follow an iterative protocol. An iteration step consists of three parts (see Fig.~\ref{fig:GradBEA_graph}a): First, a standard DMRG simulation is performed using the current matrix elements $t^{(s)}_{ij},v^{(ss')}_{ijkl}$. From the approximate ground state $\ket{\Psi_0^{\mathrm{MPS}}}$, we determine the gradient $\frac{\partial}{\partial \vec{\theta}}\calB_\mu$ over the vector $\vec{\theta}$ that parametrizes unitaries $U^{(s)}$ as we detail below. Finally, we move straight antiparallel to the gradient until $\calB_\mu$ is minimized, and capture the unitary $U^{(s)}_n$ with which we update the matrix elements $t^{(s)}$ and $v^{(ss')}$ to complete the iteration step. The total unitary after $n_\mathrm{tot}$ iterations is the product $U^{(s)} = U^{(s)}_{n_\mathrm{tot}} U^{(s)}_{n_\mathrm{tot}-1} \dots U^{(s)}_1$.

Within this work, we consider only time-reversal symmetric models, where $t^{(s)}$ and $v^{(ss')}$ can be kept real if allowing rotations only from the orthogonal group, i.e.,  $U^{(s)} \in SO(N)$. Infinitesimal rotations are then parametrized as 
\begin{equation}
    U_{ij}^{(s)} = \delta_{ij} + \sum_{(kl)} \Delta \theta^{(s)}_{(kl)}X^{(kl)}_{ij} \; ,
\end{equation}
where $\delta_{ij}$ is the Kronecker-delta, the summation goes for mode pairs $(kl)$ with $k < l$, $\Delta \theta^{(s)}_{(kl)}$ is an infinitesimal angle and 
$X^{(kl)}_{ij} = \delta_{ki}\delta_{lj} - \delta_{kj}\delta_{li}$ is the antisymmetric generator of the rotation between modes $k$ and $l$. Such an infinitesimal rotation translates to the following many-body rotation of the state vector
\begin{equation}
    \ket{\tilde{\Psi}_0} = \ket{\Psi_0}+ \sum_{(kl)} \Delta \theta^{(s)}_{(kl)}\hat{T}^{(s)}_{(kl)} \ket{\Psi_0}\; ,
\end{equation}
with $\hat{T}^{(s)}_{(kl)} = c^\dag_{ks} c_{ls} - c^\dag_{ls} c_{ks}$, the antisymmetric generator of mode rotation in the Fock space. We determine the derivatives of the individual Schmidt weights $p_\alpha=\lambda_\alpha^2$ at every bond. For the subsystems $A$ and $B$ introduced above, the reduced density matrix in subsystem $A$ before rotation is
\begin{equation}
    \hat{\rho}_A = \mathrm{Tr}_B \ket{\Psi_0} \bra{\Psi_0} = \sum_{\alpha}p_\alpha \ket{\alpha}_A \bra{\alpha}_A \;.
\end{equation}
After an infinitesimal rotation by $\Delta\theta^{(s)}_{(kl)}$, keeping only linear terms, the reduced density matrix transforms to
\begin{eqnarray}
    \hat{\rho}'_A &=& \mathrm{Tr}_B \left\lbrace \ket{\Psi_0} \bra{\Psi_0} + \Delta \theta^{(s)}_{(kl)}\left(\ket{\Xi} \bra{\Psi_0}+\ket{\Psi_0} \bra{\Xi} \right)  \right\rbrace \nonumber \\ &=& \sum_{\alpha}p_\alpha \ket{\alpha}_A \bra{\alpha}_A + \nonumber \\ & &  \Delta \theta^{(s)}_{(kl)} \sum_{\alpha \beta} (\lambda_\alpha\Xi_{\beta \alpha}^{(s,kl)*} + \Xi^{(s,kl)}_{\alpha \beta} \lambda_\beta)  \ket{\alpha}_A \bra{\beta}_A\;. \label{eq:RDM_A_transformed}
\end{eqnarray}
Here, $\ket{\Xi} = \hat{T}^{(s)}_{(kl)} \ket{\Psi_0} = \sum_{\alpha,\beta} \Xi^{(s,kl)}_{\alpha \beta} \ket{\alpha}_A \ket{\beta}_B$ has been introduced for convenience. The derivative of the eigenvalues of $\hat{\rho}'_A$ are determined from first-order non-degenerate perturbation theory~\footnote{We note that this formula should be slightly modified for degenerate Schmidt values, but in our numerical examples this was not necessary.},
\begin{equation}
\frac{\partial p_{\alpha}}{\partial \theta^{(s)}_{(kl)}} = 2 \lambda_\alpha \mathrm{Re}\left\lbrace \Xi_{\alpha\alpha}^{(s,kl)}\right\rbrace \; . \label{eq:palpha_derivative}
\end{equation}
The derivative \eqref{eq:palpha_derivative} can be efficiently calculated if the MPS is stored in the mixed canonical form usually used by DMRG (see the contraction graph in Fig.~\ref{fig:GradBEA_graph}b). Nonzero derivatives are only measured for those $(kl)$ pairs that correspond to mode rotations between the two subsystems. The derivative of the entropy for the given bipartition is
\begin{equation}
    \frac{\partial S_\mu^{(l)}(\Psi)}{\partial \theta_{ij}} = \frac{1}{1-\mu} \frac{\sum_\alpha \mu p_\alpha^{\mu-1} \frac{\partial p_\alpha}{\partial \theta_{ij}}}{\sum_\alpha p_\alpha^\mu} \;, 
\end{equation}
and the sum of these derivatives for every bond gives us the gradient element $\frac{\partial}{\partial \theta^{(s)}_{(kl)}}\calB_\mu$. 

The gradient defines the quadratic disentangler Hamiltonian
\begin{equation}
    \hat{H}_\mathrm{dis} = -i \sum_{(kl), s} \frac{\partial \mathcal{B}_\mu}{\partial \theta^{(s)}_{(kl)}} \hat{T}^{(s)}_{(kl)} \; , \label{eq:H_dis}
\end{equation}
that locally decreases $\calB_\mu$ with steepest descent. In the third iteration step, we simulate the time dependence $\exp(-i H_{\mathrm{dis}} \tau)\ket{\Psi_0}$ using the projector splitting time-dependent variational principle (PS-TDVP)~\cite{Haegeman-2016} and monitor $\calB_\mu$, which reaches a minimum at some time $\tau^*$. The optimal $N \times N$ single-particle unitary matrix of the iteration is then 
\begin{equation}
    \mat{U}_n^{(s)} = \exp \left( - \tau^*\sum_{(kl)} \frac{\partial \mathcal{B}_\mu}{\partial \theta^{(s)}_{(kl)}} \mat{X}^{(kl)} \right) \; . \label{eq:U_n}
\end{equation}
At the end of the iteration, we rotate the matrix elements $t^{(s)}$ and $v^{(ss')}$ by $\mat{U}_n^{(s)}$ and start the DMRG step of the next iteration with these. In practice, numerical performance can be improved by using a corrected search direction instead of the raw gradient in \eqref{eq:H_dis}. We have employed the Polak-Ribi{\`e}re-Polyak conjugate gradient method that uses the gradient and the search direction of the previous iteration in the correction~\cite{Polak-1969, Polyak-1969, Nocedal-2006}. The corrected search direction then replaces the raw gradient in Eq.~\eqref{eq:H_dis}, resulting in slightly better convergence near the minimum. 
\begin{figure}
    \includegraphics[width=0.48\textwidth]{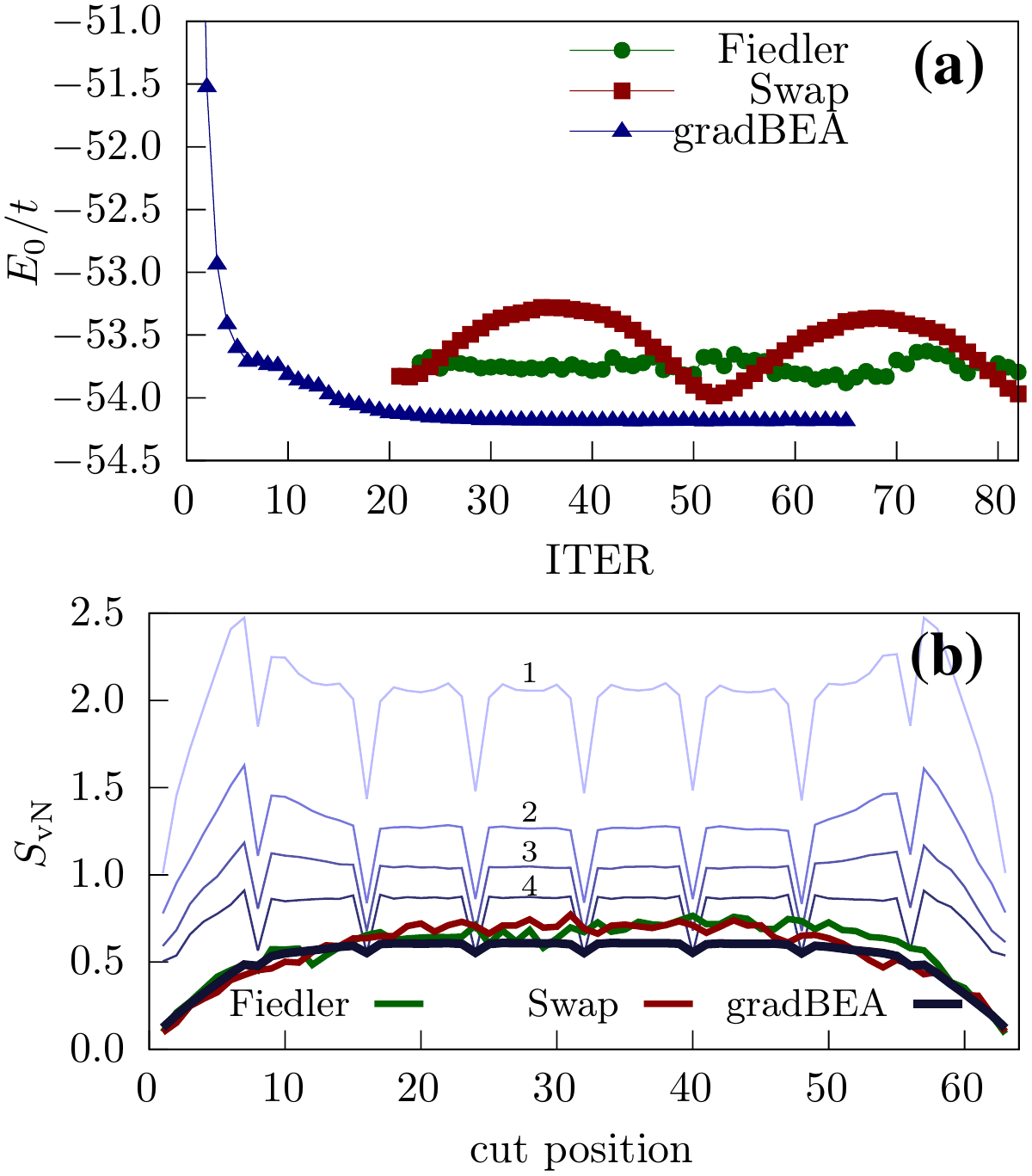}
    \caption{\textit(a) Ground state energy of the $8 \times 8$ sized Hubbard model for $U/t = 4$ as a function of mode optimization iterations at bond dimension $D=128$. Blue triangles show results using the method of this work, red squares and green circles show results using the methods of Refs.~\onlinecite{Krumnow-2016,Friesecke-2024}. For the latter two datasets the slower initial convergence is discarded by shifting the iteration number by $100$ to the left. The best (lowest) ground state energy in our work $E_0/t = -54.183$ is significantly below the best swap-gate and Fiedler-vector-based results ($E^\mathrm{swap}_0/t = -53.995$,  $E^\mathrm{Fiedler}_0/t = -5
3.882$). \label{fig:Hubbard_En_Entropy}}
\end{figure}

\emph{Numerical results\,– }First, we show results for the spinless "tt'V" model on a square lattice defined by 
\begin{equation}
\hat{H} = \left( t\sum_{\langle i,j\rangle} c^\dag_i c_j + t' \sum_{\langle i,j \rangle'}c^\dag_i c_j + \frac{V}{2} \sum_{\langle i,j \rangle} c^\dag_i c^\dag_j c_j c_i \right) + h.c.  \; ,
    \label{eq:Ham_ttV}
\end{equation}
where $\langle i,j \rangle$ and $\langle i,j \rangle'$ indicate nearest and next-nearest (diagonal) neighbors. Here the spin $s$ can take only one value, and is therefore omitted. We prescribe periodic boundary conditions in both spatial directions and use the normal site-reading order (left-to-right, top-to-bottom) initially to form the MPS. In Fig.~\ref{fig:Descent_and_Spinless}b, the ground state energy is shown as a function of mode optimization iterations for a numerically challenging parameter point close to a phase boundary ($t'/t = 0.4$, $V/t = 0.8$) on a $10 \times 10$ lattice and $D=80$, also used in Refs.~\cite{Menczer-2024a, Friesecke-2024}. The direct comparison to the Fiedler-vector- \cite{Krumnow-2016,Menczer-2024a} and systematic swap-gate-based~\cite{Friesecke-2024,Werner-2026} methods demonstrates the significant advantage of our algorithm: The initial convergence rate is an order of magnitude faster, and the converged energy is also significantly lower. Actually, the energy gets below the best value of the former methods already at the fifth iteration. We understand this rapid convergence as a consequence of the long-range nature of $\hat{H}_\mathrm{dis}$, which leads to a simultaneous and optimal mixing of modes instead of allowing only nearest-neighbor mixing. The superior final precision is a consequence of the absence of reorderings, which perturb the state in every iteration, and may lead to spurious oscillations or systematic errors near the minimum.    

The second example for which we demonstrate our method is the Hubbard model on a two-dimensional square lattice,
\begin{equation}
\hat{H} =  -t \sum_{s \in \lbrace \uparrow, \downarrow \rbrace} \sum_{\langle i,j\rangle}  \left( c^\dag_{is} c_{js} + h.c. \right) + U \sum_{i} c^\dag_{i \uparrow} c^\dag_{i \downarrow} c_{i \downarrow} c_{i \uparrow}  \label{eq:Ham_Hubbard} \; .
\end{equation}
We impose again periodic boundary conditions and initialize the MPS in the normal site-reading order. Fig.~\ref{fig:Hubbard_En_Entropy}a shows the ground state energy as a function of mode optimization iteration for $U/t = 4$ in an $8 \times 8$ lattice with bond dimension $D=128$. Results are also shown for the Fiedler-vector- and swap-gate-based methods, but for these the slow initial convergence (the first 120 iterations using the Fiedler method) is not shown for readability. The gradient-based method shows rapid convergence (the first data point $E/t = -47.203$ is outside the figure boundaries) and saturates at around $E_0/t = -54.183$. The other two methods show strong fluctuations, and their best energies are significantly higher. While the Fiedler-vector-based heuristic reorderings lead to rather random fluctuations, the systematic swap-gate-based reorderings induce regular oscillations of period $N/2$. In contrast, our algorithm is free of reorderings, which is reflected in the smooth convergence. In Fig.~\ref{fig:Hubbard_En_Entropy}b we show profiles of the von Neumann entropy $S_{\mu=1}^{(i)}$ for the first four iterations together with the optimal profiles of our method and the methods of Refs.~\cite{Krumnow-2016,Friesecke-2024}. We observe a much faster initial decrease in entropy than those shown in Refs.~\cite{Menczer-2024a,Friesecke-2024, Werner-2026}. However, we note that the entropy dips caused by the initial ordering of sites are only slowly decaying and remain slightly visible even in the last iteration. Nevertheless, the obtained profile is more regular and is significantly below the best profiles of the other two methods.  

Finally, our analysis for a difficult multireference quantum chemical system~\cite{ibm-web}, namely the Fe$_4$S$_4$ cluster subject of previous benchmark studies ~\cite{Friesecke-2024,Li-2025,Zhai-2026,Legeza-2026}, also reveal the superiority of gradBEA as reflected by the big improvement in the ground state energy using only a very limited bond dimension $D=128$: for energetic ordering we get $E_0=-327.159$, for optimized ordering $E_0=-327.182$, for Fiedler vector based mode optimization $E_0=-327.206$, for swap gate based mode optimization $E_0=-327.208$, and finally for gradBEA $E_0=-327.222$. We note that our gradBEA-based protocol, with a drastically truncated bond dimension, reduces the error significantly with respect to energy values obtained by large-scale DMRG, i.e., $E_0=-327.239$ using $D_{SU(2)}=4096$~\cite{Sharma-2014} and
$E_0=-327.244$ with $D_{SU(2)}\simeq 12000-18000$~\cite{Zhai-2026,Legeza-2026} without extrapolations.

\emph{Conclusion\,– } We constructed an algorithm that allows for efficient and robust optimization of single-particle modes and leads to optimal matrix product state description of the ground states of interacting fermionic models. The method relies on the steepest descent algorithm in which the gradient of an entanglement-based global cost function, the block entropy area $\calB_\mu$, is used to define an effective non-interacting disentangler Hamiltonian that generates the single-particle rotations.  
The method shows rapid convergence towards the optimum, where the initial convergence rate is an order of magnitude faster if compared to former methods using local (nearest-neighbor) rotations and reorderings. We attribute this rapidness to the fact that the effective Hamiltonian contains hopping terms of arbitrary distance. As our method is free of reorderings,  whose effect is especially destructive close to optimum, our optimal mode set found by our algorithm is also superior to those found by former algorithms. This superiority is reflected both in the lower value of ground state energy at the same MPS bond dimension and also in the more regular and lower entanglement entropy profile. The matrix product operator form of the non-interacting effective disentangler Hamiltonian has a low bond dimension $D_\mathrm{dis}^\mathrm{MPO} \sim \mathcal{O}(N)$, therefore the numerical requirements of the dynamical simulation via PS-TDVP remain small even if the required number of sweeps exceeds those of the ground state DMRG steps using the true, interacting Hamiltonian with $D^\mathrm{MPO} \sim \mathcal{O}(N^2)$. In this work, the method has been demonstrated for lattice models of condensed matter physics and also for the Fe${_4}$S${_4}$ cluster, showing its applicability in quantum chemistry problems~\cite{White-1999, Chan-2011, Szalay-2015, Baiardi-2020}, but the general long-ranged form \eqref{eq:Hamiltonian_spinful} of the Hamiltonian makes it also straightforwardly generalizable for nuclear shell~\cite{Dukelsky-2004, Legeza-2015, Tichai-2023} models. 
Furthermore, generalizing correlation-based optimal compression of tensor networks can also be useful in fields like neural networks or numerical analysis that are only loosely connected to quantum many-body systems.   
While the previous algorithmic solution via swap gate layers allows DMRG and mode optimization to be carried out entirely via two-qubit gates, thus the connection to quantum computing is more direct, the various favorable features of the non-local gradBEA, however, compensate for this loss on classical computers.

\emph{Acknowledgments \,–  }
This work has been supported by the Hungarian National Research, Development and Innovation Office (NKFIH) through Grant No. STARTING-152628.
\"O.L. acknowledges financial support
by the Hans Fischer Senior Fellowship programme funded by the Technical University of Munich – Institute for Advanced Study and by
the Center for Scalable and Predictive methods for Excitation and Correlated phenomena (SPEC), funded as part of the Computational Chemical Sciences Program  FWP 70942 by the U.S. Department of Energy (DOE), Office of Science, Office of Basic Energy Sciences, Division of Chemical Sciences, Geosciences, and Biosciences at Pacific Northwest National Laboratory.
K.K. has also been supported by the Janos
Bolyai Research Scholarship of the Hungarian Academy of Sciences.


\begin{thebibliography}{64}%
\makeatletter
\providecommand \@ifxundefined [1]{%
 \@ifx{#1\undefined}
}%
\providecommand \@ifnum [1]{%
 \ifnum #1\expandafter \@firstoftwo
 \else \expandafter \@secondoftwo
 \fi
}%
\providecommand \@ifx [1]{%
 \ifx #1\expandafter \@firstoftwo
 \else \expandafter \@secondoftwo
 \fi
}%
\providecommand \natexlab [1]{#1}%
\providecommand \enquote  [1]{``#1''}%
\providecommand \bibnamefont  [1]{#1}%
\providecommand \bibfnamefont [1]{#1}%
\providecommand \citenamefont [1]{#1}%
\providecommand \href@noop [0]{\@secondoftwo}%
\providecommand \href [0]{\begingroup \@sanitize@url \@href}%
\providecommand \@href[1]{\@@startlink{#1}\@@href}%
\providecommand \@@href[1]{\endgroup#1\@@endlink}%
\providecommand \@sanitize@url [0]{\catcode `\\12\catcode `\$12\catcode
  `\&12\catcode `\#12\catcode `\^12\catcode `\_12\catcode `\%12\relax}%
\providecommand \@@startlink[1]{}%
\providecommand \@@endlink[0]{}%
\providecommand \url  [0]{\begingroup\@sanitize@url \@url }%
\providecommand \@url [1]{\endgroup\@href {#1}{\urlprefix }}%
\providecommand \urlprefix  [0]{URL }%
\providecommand \Eprint [0]{\href }%
\providecommand \doibase [0]{https://doi.org/}%
\providecommand \selectlanguage [0]{\@gobble}%
\providecommand \bibinfo  [0]{\@secondoftwo}%
\providecommand \bibfield  [0]{\@secondoftwo}%
\providecommand \translation [1]{[#1]}%
\providecommand \BibitemOpen [0]{}%
\providecommand \bibitemStop [0]{}%
\providecommand \bibitemNoStop [0]{.\EOS\space}%
\providecommand \EOS [0]{\spacefactor3000\relax}%
\providecommand \BibitemShut  [1]{\csname bibitem#1\endcsname}%
\let\auto@bib@innerbib\@empty
\bibitem [{\citenamefont {Fetter}\ and\ \citenamefont
  {Walecka}(2012)}]{Fetter-2012}%
  \BibitemOpen
  \bibfield  {author} {\bibinfo {author} {\bibfnamefont {A.~L.}\ \bibnamefont
  {Fetter}}\ and\ \bibinfo {author} {\bibfnamefont {J.~D.}\ \bibnamefont
  {Walecka}},\ }\href@noop {} {\emph {\bibinfo {title} {Quantum theory of
  many-particle systems}}}\ (\bibinfo  {publisher} {Courier Corporation},\
  \bibinfo {year} {2012})\BibitemShut {NoStop}%
\bibitem [{\citenamefont {Negele}\ and\ \citenamefont
  {Orland}(2018)}]{Negele-2018}%
  \BibitemOpen
  \bibfield  {author} {\bibinfo {author} {\bibfnamefont {J.~W.}\ \bibnamefont
  {Negele}}\ and\ \bibinfo {author} {\bibfnamefont {H.}~\bibnamefont
  {Orland}},\ }\href@noop {} {\emph {\bibinfo {title} {Quantum many-particle
  systems}}}\ (\bibinfo  {publisher} {CRC Press},\ \bibinfo {year}
  {2018})\BibitemShut {NoStop}%
\bibitem [{\citenamefont {Schollw{\"o}ck}(2005)}]{Schollwock-2005}%
  \BibitemOpen
  \bibfield  {author} {\bibinfo {author} {\bibfnamefont {U.}~\bibnamefont
  {Schollw{\"o}ck}},\ }\href {https://doi.org/10.1103/RevModPhys.77.259}
  {\bibfield  {journal} {\bibinfo  {journal} {Reviews of Modern Physics}\
  }\textbf {\bibinfo {volume} {77}},\ \bibinfo {pages} {259} (\bibinfo {year}
  {2005})}\BibitemShut {NoStop}%
\bibitem [{\citenamefont {Foulkes}\ \emph {et~al.}(2001)\citenamefont
  {Foulkes}, \citenamefont {Mitas}, \citenamefont {Needs},\ and\ \citenamefont
  {Rajagopal}}]{Foulkes-2001}%
  \BibitemOpen
  \bibfield  {author} {\bibinfo {author} {\bibfnamefont {W.~M.}\ \bibnamefont
  {Foulkes}}, \bibinfo {author} {\bibfnamefont {L.}~\bibnamefont {Mitas}},
  \bibinfo {author} {\bibfnamefont {R.}~\bibnamefont {Needs}},\ and\ \bibinfo
  {author} {\bibfnamefont {G.}~\bibnamefont {Rajagopal}},\ }\href
  {https://doi.org/10.1103/RevModPhys.73.33} {\bibfield  {journal} {\bibinfo
  {journal} {Reviews of Modern Physics}\ }\textbf {\bibinfo {volume} {73}},\
  \bibinfo {pages} {33} (\bibinfo {year} {2001})}\BibitemShut {NoStop}%
\bibitem [{\citenamefont {Held}(2007)}]{Held-2007}%
  \BibitemOpen
  \bibfield  {author} {\bibinfo {author} {\bibfnamefont {K.}~\bibnamefont
  {Held}},\ }\href {https://doi.org/10.1080/00018730701619647} {\bibfield
  {journal} {\bibinfo  {journal} {Advances in Physics}\ }\textbf {\bibinfo
  {volume} {56}},\ \bibinfo {pages} {829} (\bibinfo {year} {2007})}\BibitemShut
  {NoStop}%
\bibitem [{\citenamefont {Georges}\ \emph {et~al.}(1996)\citenamefont
  {Georges}, \citenamefont {Kotliar}, \citenamefont {Krauth},\ and\
  \citenamefont {Rozenberg}}]{Georges-1996}%
  \BibitemOpen
  \bibfield  {author} {\bibinfo {author} {\bibfnamefont {A.}~\bibnamefont
  {Georges}}, \bibinfo {author} {\bibfnamefont {G.}~\bibnamefont {Kotliar}},
  \bibinfo {author} {\bibfnamefont {W.}~\bibnamefont {Krauth}},\ and\ \bibinfo
  {author} {\bibfnamefont {M.~J.}\ \bibnamefont {Rozenberg}},\ }\href
  {https://doi.org/10.1103/RevModPhys.68.13} {\bibfield  {journal} {\bibinfo
  {journal} {Reviews of Modern Physics}\ }\textbf {\bibinfo {volume} {68}},\
  \bibinfo {pages} {13} (\bibinfo {year} {1996})}\BibitemShut {NoStop}%
\bibitem [{\citenamefont {Bartlett}\ and\ \citenamefont
  {Musia{\l}}(2007)}]{Bartlett-2007}%
  \BibitemOpen
  \bibfield  {author} {\bibinfo {author} {\bibfnamefont {R.~J.}\ \bibnamefont
  {Bartlett}}\ and\ \bibinfo {author} {\bibfnamefont {M.}~\bibnamefont
  {Musia{\l}}},\ }\href {https://doi.org/10.1103/RevModPhys.79.291} {\bibfield
  {journal} {\bibinfo  {journal} {Reviews of Modern Physics}\ }\textbf
  {\bibinfo {volume} {79}},\ \bibinfo {pages} {291} (\bibinfo {year}
  {2007})}\BibitemShut {NoStop}%
\bibitem [{\citenamefont {Metzner}\ \emph {et~al.}(2012)\citenamefont
  {Metzner}, \citenamefont {Salmhofer}, \citenamefont {Honerkamp},
  \citenamefont {Meden},\ and\ \citenamefont {Sch{\"o}nhammer}}]{Metzner-2012}%
  \BibitemOpen
  \bibfield  {author} {\bibinfo {author} {\bibfnamefont {W.}~\bibnamefont
  {Metzner}}, \bibinfo {author} {\bibfnamefont {M.}~\bibnamefont {Salmhofer}},
  \bibinfo {author} {\bibfnamefont {C.}~\bibnamefont {Honerkamp}}, \bibinfo
  {author} {\bibfnamefont {V.}~\bibnamefont {Meden}},\ and\ \bibinfo {author}
  {\bibfnamefont {K.}~\bibnamefont {Sch{\"o}nhammer}},\ }\href
  {https://doi.org/10.1103/RevModPhys.84.299} {\bibfield  {journal} {\bibinfo
  {journal} {Reviews of Modern Physics}\ }\textbf {\bibinfo {volume} {84}},\
  \bibinfo {pages} {299} (\bibinfo {year} {2012})}\BibitemShut {NoStop}%
\bibitem [{\citenamefont {Carleo}\ and\ \citenamefont
  {Troyer}(2017)}]{Carleo-2017}%
  \BibitemOpen
  \bibfield  {author} {\bibinfo {author} {\bibfnamefont {G.}~\bibnamefont
  {Carleo}}\ and\ \bibinfo {author} {\bibfnamefont {M.}~\bibnamefont
  {Troyer}},\ }\href {https://doi.org/10.1126/science.aag2302} {\bibfield
  {journal} {\bibinfo  {journal} {Science}\ }\textbf {\bibinfo {volume}
  {355}},\ \bibinfo {pages} {602} (\bibinfo {year} {2017})}\BibitemShut
  {NoStop}%
\bibitem [{\citenamefont {White}(1992)}]{White-1992}%
  \BibitemOpen
  \bibfield  {author} {\bibinfo {author} {\bibfnamefont {S.~R.}\ \bibnamefont
  {White}},\ }\href {https://doi.org/10.1103/PhysRevLett.68.3487} {\bibfield
  {journal} {\bibinfo  {journal} {Physical Review Letters}\ }\textbf {\bibinfo
  {volume} {69}},\ \bibinfo {pages} {2863} (\bibinfo {year}
  {1992})}\BibitemShut {NoStop}%
\bibitem [{\citenamefont {Schollw{\"o}ck}(2011)}]{Schollwock-2011}%
  \BibitemOpen
  \bibfield  {author} {\bibinfo {author} {\bibfnamefont {U.}~\bibnamefont
  {Schollw{\"o}ck}},\ }\href {https://doi.org/10.1016/j.aop.2010.09.012}
  {\bibfield  {journal} {\bibinfo  {journal} {Annals of physics}\ }\textbf
  {\bibinfo {volume} {326}},\ \bibinfo {pages} {96} (\bibinfo {year}
  {2011})}\BibitemShut {NoStop}%
\bibitem [{\citenamefont {Or{\'u}s}(2014)}]{Orus-2014}%
  \BibitemOpen
  \bibfield  {author} {\bibinfo {author} {\bibfnamefont {R.}~\bibnamefont
  {Or{\'u}s}},\ }\href {https://doi.org/10.1016/j.aop.2014.06.013} {\bibfield
  {journal} {\bibinfo  {journal} {Annals of Physics}\ }\textbf {\bibinfo
  {volume} {349}},\ \bibinfo {pages} {117} (\bibinfo {year}
  {2014})}\BibitemShut {NoStop}%
\bibitem [{\citenamefont {Szalay}\ \emph {et~al.}(2015)\citenamefont {Szalay},
  \citenamefont {Pfeffer}, \citenamefont {Murg}, \citenamefont {Barcza},
  \citenamefont {Verstraete}, \citenamefont {Schneider},\ and\ \citenamefont
  {Legeza}}]{Szalay-2015}%
  \BibitemOpen
  \bibfield  {author} {\bibinfo {author} {\bibfnamefont {S.}~\bibnamefont
  {Szalay}}, \bibinfo {author} {\bibfnamefont {M.}~\bibnamefont {Pfeffer}},
  \bibinfo {author} {\bibfnamefont {V.}~\bibnamefont {Murg}}, \bibinfo {author}
  {\bibfnamefont {G.}~\bibnamefont {Barcza}}, \bibinfo {author} {\bibfnamefont
  {F.}~\bibnamefont {Verstraete}}, \bibinfo {author} {\bibfnamefont
  {R.}~\bibnamefont {Schneider}},\ and\ \bibinfo {author} {\bibfnamefont
  {{\"O}.}~\bibnamefont {Legeza}},\ }\href {https://doi.org/10.1002/qua.24898}
  {\bibfield  {journal} {\bibinfo  {journal} {International Journal of Quantum
  Chemistry}\ }\textbf {\bibinfo {volume} {115}},\ \bibinfo {pages} {1342}
  (\bibinfo {year} {2015})}\BibitemShut {NoStop}%
\bibitem [{\citenamefont {Cirac}\ \emph {et~al.}(2021)\citenamefont {Cirac},
  \citenamefont {Perez-Garcia}, \citenamefont {Schuch},\ and\ \citenamefont
  {Verstraete}}]{Cirac-2021}%
  \BibitemOpen
  \bibfield  {author} {\bibinfo {author} {\bibfnamefont {J.~I.}\ \bibnamefont
  {Cirac}}, \bibinfo {author} {\bibfnamefont {D.}~\bibnamefont {Perez-Garcia}},
  \bibinfo {author} {\bibfnamefont {N.}~\bibnamefont {Schuch}},\ and\ \bibinfo
  {author} {\bibfnamefont {F.}~\bibnamefont {Verstraete}},\ }\href
  {https://doi.org/10.1103/RevModPhys.93.045003} {\bibfield  {journal}
  {\bibinfo  {journal} {Reviews of Modern Physics}\ }\textbf {\bibinfo {volume}
  {93}},\ \bibinfo {pages} {045003} (\bibinfo {year} {2021})}\BibitemShut
  {NoStop}%
\bibitem [{\citenamefont {Verstraete}\ \emph {et~al.}(2023)\citenamefont
  {Verstraete}, \citenamefont {Nishino}, \citenamefont {Schollw{\"o}ck},
  \citenamefont {Ba{\~n}uls}, \citenamefont {Chan},\ and\ \citenamefont
  {Stoudenmire}}]{Verstraete-2023}%
  \BibitemOpen
  \bibfield  {author} {\bibinfo {author} {\bibfnamefont {F.}~\bibnamefont
  {Verstraete}}, \bibinfo {author} {\bibfnamefont {T.}~\bibnamefont {Nishino}},
  \bibinfo {author} {\bibfnamefont {U.}~\bibnamefont {Schollw{\"o}ck}},
  \bibinfo {author} {\bibfnamefont {M.~C.}\ \bibnamefont {Ba{\~n}uls}},
  \bibinfo {author} {\bibfnamefont {G.~K.}\ \bibnamefont {Chan}},\ and\
  \bibinfo {author} {\bibfnamefont {M.~E.}\ \bibnamefont {Stoudenmire}},\
  }\href {https://doi.org/10.1038/s42254-023-00572-5} {\bibfield  {journal}
  {\bibinfo  {journal} {Nature Reviews Physics}\ ,\ \bibinfo {pages} {1}}
  (\bibinfo {year} {2023})}\BibitemShut {NoStop}%
\bibitem [{\citenamefont {Schuch}\ \emph {et~al.}(2008)\citenamefont {Schuch},
  \citenamefont {Wolf}, \citenamefont {Verstraete},\ and\ \citenamefont
  {Cirac}}]{Schuch-2008}%
  \BibitemOpen
  \bibfield  {author} {\bibinfo {author} {\bibfnamefont {N.}~\bibnamefont
  {Schuch}}, \bibinfo {author} {\bibfnamefont {M.~M.}\ \bibnamefont {Wolf}},
  \bibinfo {author} {\bibfnamefont {F.}~\bibnamefont {Verstraete}},\ and\
  \bibinfo {author} {\bibfnamefont {J.~I.}\ \bibnamefont {Cirac}},\ }\href
  {https://doi.org/10.1103/PhysRevLett.100.030504} {\bibfield  {journal}
  {\bibinfo  {journal} {Physical Review Letters}\ }\textbf {\bibinfo {volume}
  {100}},\ \bibinfo {pages} {030504} (\bibinfo {year} {2008})}\BibitemShut
  {NoStop}%
\bibitem [{\citenamefont {Eisert}\ \emph {et~al.}(2010)\citenamefont {Eisert},
  \citenamefont {Cramer},\ and\ \citenamefont {Plenio}}]{Eisert-2010}%
  \BibitemOpen
  \bibfield  {author} {\bibinfo {author} {\bibfnamefont {J.}~\bibnamefont
  {Eisert}}, \bibinfo {author} {\bibfnamefont {M.}~\bibnamefont {Cramer}},\
  and\ \bibinfo {author} {\bibfnamefont {M.~B.}\ \bibnamefont {Plenio}},\
  }\href {https://doi.org/10.1103/RevModPhys.82.277} {\bibfield  {journal}
  {\bibinfo  {journal} {Rev. Mod. Phys.}\ }\textbf {\bibinfo {volume} {82}},\
  \bibinfo {pages} {277} (\bibinfo {year} {2010})}\BibitemShut {NoStop}%
\bibitem [{\citenamefont {Stoudenmire}\ and\ \citenamefont
  {White}(2012)}]{Stoudenmire-2012}%
  \BibitemOpen
  \bibfield  {author} {\bibinfo {author} {\bibfnamefont {E.~M.}\ \bibnamefont
  {Stoudenmire}}\ and\ \bibinfo {author} {\bibfnamefont {S.~R.}\ \bibnamefont
  {White}},\ }\href {https://doi.org/10.1146/annurev-conmatphys-020911-125018}
  {\bibfield  {journal} {\bibinfo  {journal} {Annu. Rev. Condens. Matter
  Phys.}\ }\textbf {\bibinfo {volume} {3}},\ \bibinfo {pages} {111} (\bibinfo
  {year} {2012})}\BibitemShut {NoStop}%
\bibitem [{\citenamefont {Menczer}\ and\ \citenamefont
  {Legeza}(2024)}]{Menczer-2024b}%
  \BibitemOpen
  \bibfield  {author} {\bibinfo {author} {\bibfnamefont {A.}~\bibnamefont
  {Menczer}}\ and\ \bibinfo {author} {\bibfnamefont {O.}~\bibnamefont
  {Legeza}},\ }\href {https://doi.org/10.1021/acs.jctc.4c00800} {\bibfield
  {journal} {\bibinfo  {journal} {Journal of Chemical Theory and Computation}\
  }\textbf {\bibinfo {volume} {20}},\ \bibinfo {pages} {8897} (\bibinfo {year}
  {2024})}\BibitemShut {NoStop}%
\bibitem [{\citenamefont {Menczer}\ \emph
  {et~al.}(2024{\natexlab{a}})\citenamefont {Menczer}, \citenamefont {van
  Damme}, \citenamefont {Rask}, \citenamefont {Huntington}, \citenamefont
  {Hammond}, \citenamefont {Xantheas}, \citenamefont {Ganahl},\ and\
  \citenamefont {Legeza}}]{Menczer-2024c}%
  \BibitemOpen
  \bibfield  {author} {\bibinfo {author} {\bibfnamefont {A.}~\bibnamefont
  {Menczer}}, \bibinfo {author} {\bibfnamefont {M.}~\bibnamefont {van Damme}},
  \bibinfo {author} {\bibfnamefont {A.}~\bibnamefont {Rask}}, \bibinfo {author}
  {\bibfnamefont {L.}~\bibnamefont {Huntington}}, \bibinfo {author}
  {\bibfnamefont {J.}~\bibnamefont {Hammond}}, \bibinfo {author} {\bibfnamefont
  {S.~S.}\ \bibnamefont {Xantheas}}, \bibinfo {author} {\bibfnamefont
  {M.}~\bibnamefont {Ganahl}},\ and\ \bibinfo {author} {\bibfnamefont
  {O.}~\bibnamefont {Legeza}},\ }\href
  {https://doi.org/10.1021/acs.jctc.4c00903} {\bibfield  {journal} {\bibinfo
  {journal} {Journal of Chemical Theory and Computation}\ }\textbf {\bibinfo
  {volume} {20}},\ \bibinfo {pages} {8397} (\bibinfo {year}
  {2024}{\natexlab{a}})}\BibitemShut {NoStop}%
\bibitem [{\citenamefont {Menczer}\ \emph
  {et~al.}(2024{\natexlab{b}})\citenamefont {Menczer}, \citenamefont {Kap\'as},
  \citenamefont {Werner},\ and\ \citenamefont {Legeza}}]{Menczer-2024a}%
  \BibitemOpen
  \bibfield  {author} {\bibinfo {author} {\bibfnamefont {A.}~\bibnamefont
  {Menczer}}, \bibinfo {author} {\bibfnamefont {K.}~\bibnamefont {Kap\'as}},
  \bibinfo {author} {\bibfnamefont {M.~A.}\ \bibnamefont {Werner}},\ and\
  \bibinfo {author} {\bibfnamefont {{\"O}.}~\bibnamefont {Legeza}},\ }\href
  {https://doi.org/10.1103/PhysRevB.109.195148} {\bibfield  {journal} {\bibinfo
   {journal} {Phys. Rev. B}\ }\textbf {\bibinfo {volume} {109}},\ \bibinfo
  {pages} {195148} (\bibinfo {year} {2024}{\natexlab{b}})}\BibitemShut
  {NoStop}%
\bibitem [{\citenamefont {Xiang}\ \emph {et~al.}(2024)\citenamefont {Xiang},
  \citenamefont {Jia}, \citenamefont {Fang},\ and\ \citenamefont
  {Li}}]{Xiang-2024}%
  \BibitemOpen
  \bibfield  {author} {\bibinfo {author} {\bibfnamefont {C.}~\bibnamefont
  {Xiang}}, \bibinfo {author} {\bibfnamefont {W.}~\bibnamefont {Jia}}, \bibinfo
  {author} {\bibfnamefont {W.-H.}\ \bibnamefont {Fang}},\ and\ \bibinfo
  {author} {\bibfnamefont {Z.}~\bibnamefont {Li}},\ }\href
  {https://doi.org/10.1021/acs.jctc.3c01228} {\bibfield  {journal} {\bibinfo
  {journal} {Journal of Chemical Theory and Computation}\ }\textbf {\bibinfo
  {volume} {20}},\ \bibinfo {pages} {775} (\bibinfo {year} {2024})}\BibitemShut
  {NoStop}%
\bibitem [{\citenamefont {Krumnow}\ \emph {et~al.}(2016)\citenamefont
  {Krumnow}, \citenamefont {Veis}, \citenamefont {Legeza},\ and\ \citenamefont
  {Eisert}}]{Krumnow-2016}%
  \BibitemOpen
  \bibfield  {author} {\bibinfo {author} {\bibfnamefont {C.}~\bibnamefont
  {Krumnow}}, \bibinfo {author} {\bibfnamefont {L.}~\bibnamefont {Veis}},
  \bibinfo {author} {\bibfnamefont {{\"O}.}~\bibnamefont {Legeza}},\ and\
  \bibinfo {author} {\bibfnamefont {J.}~\bibnamefont {Eisert}},\ }\href
  {https://doi.org/10.1103/PhysRevLett.117.210402} {\bibfield  {journal}
  {\bibinfo  {journal} {Phys. Rev. Lett.}\ }\textbf {\bibinfo {volume} {117}},\
  \bibinfo {pages} {210402} (\bibinfo {year} {2016})}\BibitemShut {NoStop}%
\bibitem [{\citenamefont {Friesecke}\ \emph {et~al.}(2024)\citenamefont
  {Friesecke}, \citenamefont {Werner}, \citenamefont {Kapás}, \citenamefont
  {Menczer},\ and\ \citenamefont {Örs Legeza}}]{Friesecke-2024}%
  \BibitemOpen
  \bibfield  {author} {\bibinfo {author} {\bibfnamefont {G.}~\bibnamefont
  {Friesecke}}, \bibinfo {author} {\bibfnamefont {M.~A.}\ \bibnamefont
  {Werner}}, \bibinfo {author} {\bibfnamefont {K.}~\bibnamefont {Kapás}},
  \bibinfo {author} {\bibfnamefont {A.}~\bibnamefont {Menczer}},\ and\ \bibinfo
  {author} {\bibnamefont {Örs Legeza}},\ }\href
  {https://arxiv.org/abs/2406.03449} {\bibinfo {title} {Global fermionic mode
  optimization via swap gates}} (\bibinfo {year} {2024}),\ \Eprint
  {https://arxiv.org/abs/2406.03449} {arXiv:2406.03449 [cond-mat.str-el]}
  \BibitemShut {NoStop}%
\bibitem [{\citenamefont {Xiang}(1996)}]{Xiang-1996}%
  \BibitemOpen
  \bibfield  {author} {\bibinfo {author} {\bibfnamefont {T.}~\bibnamefont
  {Xiang}},\ }\href {https://doi.org/10.1103/PhysRevB.53.R10445} {\bibfield
  {journal} {\bibinfo  {journal} {Phys. Rev. B}\ }\textbf {\bibinfo {volume}
  {53}},\ \bibinfo {pages} {R10445} (\bibinfo {year} {1996})}\BibitemShut
  {NoStop}%
\bibitem [{\citenamefont {Rissler}\ \emph {et~al.}(2006)\citenamefont
  {Rissler}, \citenamefont {Noack},\ and\ \citenamefont
  {White}}]{Rissler-2006}%
  \BibitemOpen
  \bibfield  {author} {\bibinfo {author} {\bibfnamefont {J.}~\bibnamefont
  {Rissler}}, \bibinfo {author} {\bibfnamefont {R.~M.}\ \bibnamefont {Noack}},\
  and\ \bibinfo {author} {\bibfnamefont {S.~R.}\ \bibnamefont {White}},\ }\href
  {https://doi.org/http://dx.doi.org/10.1016/j.chemphys.2005.10.018} {\bibfield
   {journal} {\bibinfo  {journal} {Chemical Physics}\ }\textbf {\bibinfo
  {volume} {323}},\ \bibinfo {pages} {519 } (\bibinfo {year}
  {2006})}\BibitemShut {NoStop}%
\bibitem [{\citenamefont {Legeza}\ \emph {et~al.}(2006)\citenamefont {Legeza},
  \citenamefont {Gebhard},\ and\ \citenamefont {Rissler}}]{Legeza-2006b}%
  \BibitemOpen
  \bibfield  {author} {\bibinfo {author} {\bibfnamefont {{\"O}.}~\bibnamefont
  {Legeza}}, \bibinfo {author} {\bibfnamefont {F.}~\bibnamefont {Gebhard}},\
  and\ \bibinfo {author} {\bibfnamefont {J.}~\bibnamefont {Rissler}},\ }\href
  {https://doi.org/10.1103/PhysRevB.74.195112} {\bibfield  {journal} {\bibinfo
  {journal} {Phys. Rev. B}\ }\textbf {\bibinfo {volume} {74}},\ \bibinfo
  {pages} {195112} (\bibinfo {year} {2006})}\BibitemShut {NoStop}%
\bibitem [{\citenamefont {Murg}\ \emph {et~al.}(2010)\citenamefont {Murg},
  \citenamefont {Verstraete}, \citenamefont {Legeza},\ and\ \citenamefont
  {Noack}}]{Murg-2010a}%
  \BibitemOpen
  \bibfield  {author} {\bibinfo {author} {\bibfnamefont {V.}~\bibnamefont
  {Murg}}, \bibinfo {author} {\bibfnamefont {F.}~\bibnamefont {Verstraete}},
  \bibinfo {author} {\bibfnamefont {{\"O}.}~\bibnamefont {Legeza}},\ and\
  \bibinfo {author} {\bibfnamefont {R.~M.}\ \bibnamefont {Noack}},\ }\href
  {https://doi.org/10.1103/PhysRevB.82.205105} {\bibfield  {journal} {\bibinfo
  {journal} {Phys. Rev. B}\ }\textbf {\bibinfo {volume} {82}},\ \bibinfo
  {pages} {205105} (\bibinfo {year} {2010})}\BibitemShut {NoStop}%
\bibitem [{\citenamefont {Olivares-Amaya}\ \emph {et~al.}(2015)\citenamefont
  {Olivares-Amaya}, \citenamefont {Hu}, \citenamefont {Nakatani}, \citenamefont
  {Sharma}, \citenamefont {Yang},\ and\ \citenamefont {Chan}}]{Olivares-2015}%
  \BibitemOpen
  \bibfield  {author} {\bibinfo {author} {\bibfnamefont {R.}~\bibnamefont
  {Olivares-Amaya}}, \bibinfo {author} {\bibfnamefont {W.}~\bibnamefont {Hu}},
  \bibinfo {author} {\bibfnamefont {N.}~\bibnamefont {Nakatani}}, \bibinfo
  {author} {\bibfnamefont {S.}~\bibnamefont {Sharma}}, \bibinfo {author}
  {\bibfnamefont {J.}~\bibnamefont {Yang}},\ and\ \bibinfo {author}
  {\bibfnamefont {G.~K.}\ \bibnamefont {Chan}},\ }\href
  {https://doi.org/10.1063/1.4905329} {\bibfield  {journal} {\bibinfo
  {journal} {The Journal of Chemical Physics}\ }\textbf {\bibinfo {volume}
  {142}} (\bibinfo {year} {2015})}\BibitemShut {NoStop}%
\bibitem [{\citenamefont {Keller}\ and\ \citenamefont
  {Reiher}(0000)}]{Keller-2014}%
  \BibitemOpen
  \bibfield  {author} {\bibinfo {author} {\bibfnamefont {S.~F.}\ \bibnamefont
  {Keller}}\ and\ \bibinfo {author} {\bibfnamefont {M.}~\bibnamefont
  {Reiher}},\ }\href {https://doi.org/doi:10.2533/chimia.2014.200} {\bibfield
  {journal} {\bibinfo  {journal} {CHIMIA International Journal for Chemistry}\
  }\textbf {\bibinfo {volume} {68}},\ \bibinfo {pages} {200} (\bibinfo {year}
  {2014-04-30T00:00:00})}\BibitemShut {NoStop}%
\bibitem [{\citenamefont {Sharma}\ \emph {et~al.}(2014)\citenamefont {Sharma},
  \citenamefont {Sivalingam}, \citenamefont {Neese},\ and\ \citenamefont
  {Chan}}]{Sharma-2014}%
  \BibitemOpen
  \bibfield  {author} {\bibinfo {author} {\bibfnamefont {S.}~\bibnamefont
  {Sharma}}, \bibinfo {author} {\bibfnamefont {K.}~\bibnamefont {Sivalingam}},
  \bibinfo {author} {\bibfnamefont {F.}~\bibnamefont {Neese}},\ and\ \bibinfo
  {author} {\bibfnamefont {G.~K.-L.}\ \bibnamefont {Chan}},\ }\href
  {https://doi.org/10.1038/nchem.2041} {\bibfield  {journal} {\bibinfo
  {journal} {Nature Chemistry}\ }\textbf {\bibinfo {volume} {6}},\ \bibinfo
  {pages} {927} (\bibinfo {year} {2014})}\BibitemShut {NoStop}%
\bibitem [{\citenamefont {Vidal}\ \emph {et~al.}(2003)\citenamefont {Vidal},
  \citenamefont {Latorre}, \citenamefont {Rico},\ and\ \citenamefont
  {Kitaev}}]{Vidal-2003}%
  \BibitemOpen
  \bibfield  {author} {\bibinfo {author} {\bibfnamefont {G.}~\bibnamefont
  {Vidal}}, \bibinfo {author} {\bibfnamefont {J.~I.}\ \bibnamefont {Latorre}},
  \bibinfo {author} {\bibfnamefont {E.}~\bibnamefont {Rico}},\ and\ \bibinfo
  {author} {\bibfnamefont {A.}~\bibnamefont {Kitaev}},\ }\href
  {https://doi.org/10.1103/PhysRevLett.90.227902} {\bibfield  {journal}
  {\bibinfo  {journal} {Physical Review Letters}\ }\textbf {\bibinfo {volume}
  {90}},\ \bibinfo {pages} {227902} (\bibinfo {year} {2003})}\BibitemShut
  {NoStop}%
\bibitem [{\citenamefont {Legeza}\ and\ \citenamefont
  {S\'olyom}(2003)}]{Legeza-2003b}%
  \BibitemOpen
  \bibfield  {author} {\bibinfo {author} {\bibfnamefont {{\"O}.}~\bibnamefont
  {Legeza}}\ and\ \bibinfo {author} {\bibfnamefont {J.}~\bibnamefont
  {S\'olyom}},\ }\href {https://doi.org/10.1103/PhysRevB.68.195116} {\bibfield
  {journal} {\bibinfo  {journal} {Phys. Rev. B}\ }\textbf {\bibinfo {volume}
  {68}},\ \bibinfo {pages} {195116} (\bibinfo {year} {2003})}\BibitemShut
  {NoStop}%
\bibitem [{\citenamefont {Verstraete}\ and\ \citenamefont
  {Cirac}(2006)}]{Verstraete-2006}%
  \BibitemOpen
  \bibfield  {author} {\bibinfo {author} {\bibfnamefont {F.}~\bibnamefont
  {Verstraete}}\ and\ \bibinfo {author} {\bibfnamefont {J.~I.}\ \bibnamefont
  {Cirac}},\ }\href {https://doi.org/10.1103/PhysRevB.73.094423} {\bibfield
  {journal} {\bibinfo  {journal} {Phys. Rev. B}\ }\textbf {\bibinfo {volume}
  {73}},\ \bibinfo {pages} {094423} (\bibinfo {year} {2006})}\BibitemShut
  {NoStop}%
\bibitem [{\citenamefont {Barcza}\ \emph {et~al.}(2011)\citenamefont {Barcza},
  \citenamefont {Legeza}, \citenamefont {Marti},\ and\ \citenamefont
  {Reiher}}]{Barcza-2011}%
  \BibitemOpen
  \bibfield  {author} {\bibinfo {author} {\bibfnamefont {G.}~\bibnamefont
  {Barcza}}, \bibinfo {author} {\bibfnamefont {{\"O}.}~\bibnamefont {Legeza}},
  \bibinfo {author} {\bibfnamefont {K.~H.}\ \bibnamefont {Marti}},\ and\
  \bibinfo {author} {\bibfnamefont {M.}~\bibnamefont {Reiher}},\ }\href
  {https://doi.org/10.1103/PhysRevA.83.012508} {\bibfield  {journal} {\bibinfo
  {journal} {Physical Review A}\ }\textbf {\bibinfo {volume} {83}},\ \bibinfo
  {pages} {012508} (\bibinfo {year} {2011})}\BibitemShut {NoStop}%
\bibitem [{\citenamefont {Krumnow}\ \emph {et~al.}(2021)\citenamefont
  {Krumnow}, \citenamefont {Veis}, \citenamefont {Eisert},\ and\ \citenamefont
  {Legeza}}]{Krumnow-2021}%
  \BibitemOpen
  \bibfield  {author} {\bibinfo {author} {\bibfnamefont {C.}~\bibnamefont
  {Krumnow}}, \bibinfo {author} {\bibfnamefont {L.}~\bibnamefont {Veis}},
  \bibinfo {author} {\bibfnamefont {J.}~\bibnamefont {Eisert}},\ and\ \bibinfo
  {author} {\bibfnamefont {{\"O}.}~\bibnamefont {Legeza}},\ }\href
  {https://doi.org/10.1103/PhysRevB.104.075137} {\bibfield  {journal} {\bibinfo
   {journal} {Phys. Rev. B}\ }\textbf {\bibinfo {volume} {104}},\ \bibinfo
  {pages} {075137} (\bibinfo {year} {2021})}\BibitemShut {NoStop}%
\bibitem [{\citenamefont {Krumnow}\ \emph {et~al.}(2019)\citenamefont
  {Krumnow}, \citenamefont {Legeza},\ and\ \citenamefont
  {Eisert}}]{Krumnow-2019}%
  \BibitemOpen
  \bibfield  {author} {\bibinfo {author} {\bibfnamefont {C.}~\bibnamefont
  {Krumnow}}, \bibinfo {author} {\bibfnamefont {{\"O}.}~\bibnamefont
  {Legeza}},\ and\ \bibinfo {author} {\bibfnamefont {J.}~\bibnamefont
  {Eisert}},\ }\href {https://arxiv.org/abs/1904.11999} {\bibfield  {journal}
  {\bibinfo  {journal} {arXiv [cond-mat.stat-mech]}\ ,\ \bibinfo {pages}
  {1904.11999}} (\bibinfo {year} {2019})}\BibitemShut {NoStop}%
\bibitem [{\citenamefont {Petrov}\ \emph {et~al.}(2024)\citenamefont {Petrov},
  \citenamefont {Ganyecz}, \citenamefont {Benedek}, \citenamefont {Olasz},
  \citenamefont {Barcza},\ and\ \citenamefont {Legeza}}]{Petrov-2023}%
  \BibitemOpen
  \bibfield  {author} {\bibinfo {author} {\bibfnamefont {K.}~\bibnamefont
  {Petrov}}, \bibinfo {author} {\bibfnamefont {A.}~\bibnamefont {Ganyecz}},
  \bibinfo {author} {\bibfnamefont {Z.}~\bibnamefont {Benedek}}, \bibinfo
  {author} {\bibfnamefont {A.}~\bibnamefont {Olasz}}, \bibinfo {author}
  {\bibfnamefont {G.}~\bibnamefont {Barcza}},\ and\ \bibinfo {author}
  {\bibfnamefont {{\"O}.}~\bibnamefont {Legeza}},\ }in\ \href
  {https://doi.org/10.1007/978-3-031-52078-5_9} {\emph {\bibinfo {booktitle}
  {Advances in Methods and Applications of Quantum Systems in Chemistry,
  Physics, and Biology Selected Proceed- ings of QSCP-XXV Conference (Torun,
  Poland, June 2022)}}},\ \bibinfo {editor} {edited by\ \bibinfo {editor}
  {\bibfnamefont {I.}~\bibnamefont {Grabowski}}, \bibinfo {editor}
  {\bibfnamefont {K.}~\bibnamefont {S\l{}owik}}, \bibinfo {editor}
  {\bibfnamefont {J.}~\bibnamefont {Maruani}},\ and\ \bibinfo {editor}
  {\bibfnamefont {E.~J.}\ \bibnamefont {Br{\"a}ndas}}}\ (\bibinfo  {publisher}
  {Springer, Cham},\ \bibinfo {year} {2024})\BibitemShut {NoStop}%
\bibitem [{\citenamefont {M{\'a}t{\'e}}\ \emph {et~al.}(2023)\citenamefont
  {M{\'a}t{\'e}}, \citenamefont {Petrov}, \citenamefont {Szalay},\ and\
  \citenamefont {Legeza}}]{Mate-2023}%
  \BibitemOpen
  \bibfield  {author} {\bibinfo {author} {\bibfnamefont {M.}~\bibnamefont
  {M{\'a}t{\'e}}}, \bibinfo {author} {\bibfnamefont {K.}~\bibnamefont
  {Petrov}}, \bibinfo {author} {\bibfnamefont {S.}~\bibnamefont {Szalay}},\
  and\ \bibinfo {author} {\bibfnamefont {{\"O}.}~\bibnamefont {Legeza}},\
  }\href {https://doi.org/10.1007/s10910-022-01379-y} {\bibfield  {journal}
  {\bibinfo  {journal} {Journal of Mathematical Chemistry}\ }\textbf {\bibinfo
  {volume} {61}},\ \bibinfo {pages} {362} (\bibinfo {year} {2023})}\BibitemShut
  {NoStop}%
\bibitem [{\citenamefont {Moca}\ \emph {et~al.}(2020)\citenamefont {Moca},
  \citenamefont {Izumida}, \citenamefont {D{\'o}ra}, \citenamefont {Legeza},
  \citenamefont {Asb{\'o}th},\ and\ \citenamefont {Zar{\'a}nd}}]{Moca-2020}%
  \BibitemOpen
  \bibfield  {author} {\bibinfo {author} {\bibfnamefont {C.~P.}\ \bibnamefont
  {Moca}}, \bibinfo {author} {\bibfnamefont {W.}~\bibnamefont {Izumida}},
  \bibinfo {author} {\bibfnamefont {B.}~\bibnamefont {D{\'o}ra}}, \bibinfo
  {author} {\bibfnamefont {{\"O}.}~\bibnamefont {Legeza}}, \bibinfo {author}
  {\bibfnamefont {J.~K.}\ \bibnamefont {Asb{\'o}th}},\ and\ \bibinfo {author}
  {\bibfnamefont {G.}~\bibnamefont {Zar{\'a}nd}},\ }\href
  {https://doi.org/10.1103/PhysRevLett.125.056401} {\bibfield  {journal}
  {\bibinfo  {journal} {Physical Review Letters}\ }\textbf {\bibinfo {volume}
  {125}},\ \bibinfo {pages} {056401} (\bibinfo {year} {2020})}\BibitemShut
  {NoStop}%
\bibitem [{\citenamefont {Werner}\ \emph {et~al.}(2026)\citenamefont {Werner},
  \citenamefont {Menczer},\ and\ \citenamefont {Legeza}}]{Werner-2026}%
  \BibitemOpen
  \bibfield  {author} {\bibinfo {author} {\bibfnamefont {M.~A.}\ \bibnamefont
  {Werner}}, \bibinfo {author} {\bibfnamefont {A.}~\bibnamefont {Menczer}},\
  and\ \bibinfo {author} {\bibfnamefont {O.}~\bibnamefont {Legeza}},\ }\bibinfo
  {title} {Tensor network state methods and quantum information theory for
  strongly correlated molecular systems},\ in\ \href
  {https://doi.org/10.1016/bs.aiq.2025.02.001} {\emph {\bibinfo {booktitle}
  {Hungarian Quantum Chemistry: Part B - Contemporary Research}}}\ (\bibinfo
  {publisher} {Elsevier},\ \bibinfo {year} {2026})\ p.\ \bibinfo {pages}
  {115–151}\BibitemShut {NoStop}%
\bibitem [{\citenamefont {Li}(2025)}]{Li-2025}%
  \BibitemOpen
  \bibfield  {author} {\bibinfo {author} {\bibfnamefont {Z.}~\bibnamefont
  {Li}},\ }\href {https://doi.org/10.1103/bwvc-z9hz} {\bibfield  {journal}
  {\bibinfo  {journal} {Phys. Rev. Lett.}\ }\textbf {\bibinfo {volume} {135}},\
  \bibinfo {pages} {210601} (\bibinfo {year} {2025})}\BibitemShut {NoStop}%
\bibitem [{\citenamefont {Nocedal}\ and\ \citenamefont
  {Wright}(2006)}]{Nocedal-2006}%
  \BibitemOpen
  \bibfield  {author} {\bibinfo {author} {\bibfnamefont {J.}~\bibnamefont
  {Nocedal}}\ and\ \bibinfo {author} {\bibfnamefont {S.~J.}\ \bibnamefont
  {Wright}},\ }\href@noop {} {\emph {\bibinfo {title} {Numerical
  optimization}}}\ (\bibinfo  {publisher} {Springer},\ \bibinfo {year}
  {2006})\BibitemShut {NoStop}%
\bibitem [{Note1()}]{Note1}%
  \BibitemOpen
  \bibinfo {note} {We note that the form of the Hamiltonian has to be
  generalized for higher spin ($S>1/2$) models.}\BibitemShut {Stop}%
\bibitem [{\citenamefont {Roos}\ \emph {et~al.}(1980)\citenamefont {Roos},
  \citenamefont {Taylor},\ and\ \citenamefont {Sigbahn}}]{Roos-1980}%
  \BibitemOpen
  \bibfield  {author} {\bibinfo {author} {\bibfnamefont {B.~O.}\ \bibnamefont
  {Roos}}, \bibinfo {author} {\bibfnamefont {P.~R.}\ \bibnamefont {Taylor}},\
  and\ \bibinfo {author} {\bibfnamefont {P.~E.}\ \bibnamefont {Sigbahn}},\
  }\href {https://doi.org/10.1016/0301-0104(80)80045-0} {\bibfield  {journal}
  {\bibinfo  {journal} {Chemical Physics}\ }\textbf {\bibinfo {volume} {48}},\
  \bibinfo {pages} {157} (\bibinfo {year} {1980})}\BibitemShut {NoStop}%
\bibitem [{\citenamefont {Zgid}\ and\ \citenamefont
  {Nooijen}(2008)}]{Zgid-2008c}%
  \BibitemOpen
  \bibfield  {author} {\bibinfo {author} {\bibfnamefont {D.}~\bibnamefont
  {Zgid}}\ and\ \bibinfo {author} {\bibfnamefont {M.}~\bibnamefont {Nooijen}},\
  }\href {https://doi.org/http://dx.doi.org/10.1063/1.2883981} {\bibfield
  {journal} {\bibinfo  {journal} {The Journal of Chemical Physics}\ }\textbf
  {\bibinfo {volume} {128}},\ \bibinfo {eid} {144116} (\bibinfo {year}
  {2008})}\BibitemShut {NoStop}%
\bibitem [{\citenamefont {Legeza}\ \emph {et~al.}(2025)\citenamefont {Legeza},
  \citenamefont {Menczer}, \citenamefont {Ganyecz}, \citenamefont {Werner},
  \citenamefont {Kapás}, \citenamefont {Hammond}, \citenamefont {Xantheas},
  \citenamefont {Ganahl},\ and\ \citenamefont {Neese}}]{Legeza-2025}%
  \BibitemOpen
  \bibfield  {author} {\bibinfo {author} {\bibfnamefont {O.}~\bibnamefont
  {Legeza}}, \bibinfo {author} {\bibfnamefont {A.}~\bibnamefont {Menczer}},
  \bibinfo {author} {\bibfnamefont {A.}~\bibnamefont {Ganyecz}}, \bibinfo
  {author} {\bibfnamefont {M.~A.}\ \bibnamefont {Werner}}, \bibinfo {author}
  {\bibfnamefont {K.}~\bibnamefont {Kapás}}, \bibinfo {author} {\bibfnamefont
  {J.}~\bibnamefont {Hammond}}, \bibinfo {author} {\bibfnamefont {S.~S.}\
  \bibnamefont {Xantheas}}, \bibinfo {author} {\bibfnamefont {M.}~\bibnamefont
  {Ganahl}},\ and\ \bibinfo {author} {\bibfnamefont {F.}~\bibnamefont
  {Neese}},\ }\href {https://doi.org/10.1021/acs.jctc.5c00571} {\bibfield
  {journal} {\bibinfo  {journal} {Journal of Chemical Theory and Computation}\
  }\textbf {\bibinfo {volume} {21}},\ \bibinfo {pages} {6545–6558} (\bibinfo
  {year} {2025})}\BibitemShut {NoStop}%
\bibitem [{\citenamefont {{\"O}stlund}\ and\ \citenamefont
  {Rommer}(1995)}]{Ostlund-1995}%
  \BibitemOpen
  \bibfield  {author} {\bibinfo {author} {\bibfnamefont {S.}~\bibnamefont
  {{\"O}stlund}}\ and\ \bibinfo {author} {\bibfnamefont {S.}~\bibnamefont
  {Rommer}},\ }\href {https://doi.org/10.1103/PhysRevLett.75.3537} {\bibfield
  {journal} {\bibinfo  {journal} {Physical Review Letters}\ }\textbf {\bibinfo
  {volume} {75}},\ \bibinfo {pages} {3537} (\bibinfo {year}
  {1995})}\BibitemShut {NoStop}%
\bibitem [{\citenamefont {Jordan}\ and\ \citenamefont
  {Wigner}(1928)}]{Jordan-1928}%
  \BibitemOpen
  \bibfield  {author} {\bibinfo {author} {\bibfnamefont {P.}~\bibnamefont
  {Jordan}}\ and\ \bibinfo {author} {\bibfnamefont {E.}~\bibnamefont
  {Wigner}},\ }\href@noop {} {\bibfield  {journal} {\bibinfo  {journal}
  {Zeitschrift f{\"u}r Physik}\ }\textbf {\bibinfo {volume} {47}},\ \bibinfo
  {pages} {631} (\bibinfo {year} {1928})}\BibitemShut {NoStop}%
\bibitem [{\citenamefont {S{\'o}lyom}(2007)}]{Solyom-2007Book1}%
  \BibitemOpen
  \bibfield  {author} {\bibinfo {author} {\bibfnamefont {J.}~\bibnamefont
  {S{\'o}lyom}},\ }\href@noop {} {\emph {\bibinfo {title} {Fundamentals of the
  Physics of Solids: Volume 1: Structure and Dynamics}}}\ (\bibinfo
  {publisher} {Springer},\ \bibinfo {year} {2007})\BibitemShut {NoStop}%
\bibitem [{\citenamefont {Schmidt}(1907)}]{Schmidt-1907}%
  \BibitemOpen
  \bibfield  {author} {\bibinfo {author} {\bibfnamefont {E.}~\bibnamefont
  {Schmidt}},\ }\href {https://doi.org/10.1007/bf01449770} {\bibfield
  {journal} {\bibinfo  {journal} {Mathematische Annalen}\ }\textbf {\bibinfo
  {volume} {63}},\ \bibinfo {pages} {433–476} (\bibinfo {year}
  {1907})}\BibitemShut {NoStop}%
\bibitem [{Note2()}]{Note2}%
  \BibitemOpen
  \bibinfo {note} {We note that this formula should be slightly modified for
  degenerate Schmidt values, but in our numerical examples this was not
  necessary.}\BibitemShut {Stop}%
\bibitem [{\citenamefont {Haegeman}\ \emph {et~al.}(2016)\citenamefont
  {Haegeman}, \citenamefont {Lubich}, \citenamefont {Oseledets}, \citenamefont
  {Vandereycken},\ and\ \citenamefont {Verstraete}}]{Haegeman-2016}%
  \BibitemOpen
  \bibfield  {author} {\bibinfo {author} {\bibfnamefont {J.}~\bibnamefont
  {Haegeman}}, \bibinfo {author} {\bibfnamefont {C.}~\bibnamefont {Lubich}},
  \bibinfo {author} {\bibfnamefont {I.}~\bibnamefont {Oseledets}}, \bibinfo
  {author} {\bibfnamefont {B.}~\bibnamefont {Vandereycken}},\ and\ \bibinfo
  {author} {\bibfnamefont {F.}~\bibnamefont {Verstraete}},\ }\href
  {http://dx.doi.org/10.1103/PhysRevB.94.165116} {\bibfield  {journal}
  {\bibinfo  {journal} {Physical Review B}\ }\textbf {\bibinfo {volume} {94}},\
  \bibinfo {pages} {165116} (\bibinfo {year} {2016})}\BibitemShut {NoStop}%
\bibitem [{\citenamefont {Polak}\ and\ \citenamefont
  {Ribiere}(1969)}]{Polak-1969}%
  \BibitemOpen
  \bibfield  {author} {\bibinfo {author} {\bibfnamefont {E.}~\bibnamefont
  {Polak}}\ and\ \bibinfo {author} {\bibfnamefont {G.}~\bibnamefont
  {Ribiere}},\ }\href {https://doi.org/10.1051/m2an/196903R100351} {\bibfield
  {journal} {\bibinfo  {journal} {Revue Fran{\c{c}}aise d'Informatique et de
  Recherche Op{\'e}rationnelle. S{\'e}rie Rouge}\ }\textbf {\bibinfo {volume}
  {3}},\ \bibinfo {pages} {35} (\bibinfo {year} {1969})}\BibitemShut {NoStop}%
\bibitem [{\citenamefont {Polyak}(1969)}]{Polyak-1969}%
  \BibitemOpen
  \bibfield  {author} {\bibinfo {author} {\bibfnamefont {B.~T.}\ \bibnamefont
  {Polyak}},\ }\href {https://doi.org/10.1016/0041-5553(69)90035-4} {\bibfield
  {journal} {\bibinfo  {journal} {USSR Computational Mathematics and
  Mathematical Physics}\ }\textbf {\bibinfo {volume} {9}},\ \bibinfo {pages}
  {94} (\bibinfo {year} {1969})}\BibitemShut {NoStop}%
\bibitem [{ibm()}]{ibm-web}%
  \BibitemOpen
  \href@noop {} {\bibinfo {title} {Quantum advantage tracker
  \url{https://quantum-advantage-tracker.github.io/trackers/variational-problems}}}\BibitemShut
  {NoStop}%
\bibitem [{\citenamefont {Zhai}\ \emph {et~al.}(2026)\citenamefont {Zhai},
  \citenamefont {Li}, \citenamefont {Zhang}, \citenamefont {Li}, \citenamefont
  {Lee},\ and\ \citenamefont {Chan}}]{Zhai-2026}%
  \BibitemOpen
  \bibfield  {author} {\bibinfo {author} {\bibfnamefont {H.}~\bibnamefont
  {Zhai}}, \bibinfo {author} {\bibfnamefont {C.}~\bibnamefont {Li}}, \bibinfo
  {author} {\bibfnamefont {X.}~\bibnamefont {Zhang}}, \bibinfo {author}
  {\bibfnamefont {Z.}~\bibnamefont {Li}}, \bibinfo {author} {\bibfnamefont
  {S.}~\bibnamefont {Lee}},\ and\ \bibinfo {author} {\bibfnamefont {G.~K.-L.}\
  \bibnamefont {Chan}},\ }\href {https://arxiv.org/abs/2601.04621} {\bibinfo
  {title} {Classical solution of the {FeMo}-cofactor model to chemical accuracy
  and its implications}} (\bibinfo {year} {2026}),\ \Eprint
  {https://arxiv.org/abs/2601.04621} {arXiv:2601.04621 [physics.chem-ph]}
  \BibitemShut {NoStop}%
\bibitem [{\citenamefont {Legeza}\ \emph {et~al.}(2026)\citenamefont {Legeza},
  \citenamefont {Menczer}, \citenamefont {Werner}, \citenamefont {Xantheas},
  \citenamefont {Neese}, \citenamefont {Ganahl}, \citenamefont {Brower},
  \citenamefont {Bernabeu}, \citenamefont {Hammond},\ and\ \citenamefont
  {Gunnels}}]{Legeza-2026}%
  \BibitemOpen
  \bibfield  {author} {\bibinfo {author} {\bibfnamefont {{\"O}.}~\bibnamefont
  {Legeza}}, \bibinfo {author} {\bibfnamefont {A.}~\bibnamefont {Menczer}},
  \bibinfo {author} {\bibfnamefont {M.~A.}\ \bibnamefont {Werner}}, \bibinfo
  {author} {\bibfnamefont {S.~S.}\ \bibnamefont {Xantheas}}, \bibinfo {author}
  {\bibfnamefont {F.}~\bibnamefont {Neese}}, \bibinfo {author} {\bibfnamefont
  {M.}~\bibnamefont {Ganahl}}, \bibinfo {author} {\bibfnamefont
  {C.}~\bibnamefont {Brower}}, \bibinfo {author} {\bibfnamefont {S.~R.}\
  \bibnamefont {Bernabeu}}, \bibinfo {author} {\bibfnamefont {J.}~\bibnamefont
  {Hammond}},\ and\ \bibinfo {author} {\bibfnamefont {J.}~\bibnamefont
  {Gunnels}},\ }\href {https://arxiv.org/abs/2603.28648} {\bibinfo {title}
  {Hunting for quantum advantage in electronic structure calculations is a
  highly non-trivial task}} (\bibinfo {year} {2026}),\ \Eprint
  {https://arxiv.org/abs/2603.28648} {arXiv:2603.28648} \BibitemShut {NoStop}%
\bibitem [{\citenamefont {White}\ and\ \citenamefont
  {Martin}(1999)}]{White-1999}%
  \BibitemOpen
  \bibfield  {author} {\bibinfo {author} {\bibfnamefont {S.~R.}\ \bibnamefont
  {White}}\ and\ \bibinfo {author} {\bibfnamefont {R.~L.}\ \bibnamefont
  {Martin}},\ }\href {https://doi.org/10.1063/1.478295} {\bibfield  {journal}
  {\bibinfo  {journal} {The Journal of Chemical Physics}\ }\textbf {\bibinfo
  {volume} {110}},\ \bibinfo {pages} {4127} (\bibinfo {year}
  {1999})}\BibitemShut {NoStop}%
\bibitem [{\citenamefont {Chan}\ and\ \citenamefont
  {Sharma}(2011)}]{Chan-2011}%
  \BibitemOpen
  \bibfield  {author} {\bibinfo {author} {\bibfnamefont {G.~K.-L.}\
  \bibnamefont {Chan}}\ and\ \bibinfo {author} {\bibfnamefont {S.}~\bibnamefont
  {Sharma}},\ }\href {https://doi.org/10.1146/annurev-physchem-032210-103338}
  {\bibfield  {journal} {\bibinfo  {journal} {Annual Review of Physical
  Chemistry}\ }\textbf {\bibinfo {volume} {62}},\ \bibinfo {pages} {465}
  (\bibinfo {year} {2011})}\BibitemShut {NoStop}%
\bibitem [{\citenamefont {Baiardi}\ and\ \citenamefont
  {Reiher}(2020)}]{Baiardi-2020}%
  \BibitemOpen
  \bibfield  {author} {\bibinfo {author} {\bibfnamefont {A.}~\bibnamefont
  {Baiardi}}\ and\ \bibinfo {author} {\bibfnamefont {M.}~\bibnamefont
  {Reiher}},\ }\href {https://doi.org/10.1063/1.5129672} {\bibfield  {journal}
  {\bibinfo  {journal} {The Journal of Chemical Physics}\ }\textbf {\bibinfo
  {volume} {152}},\ \bibinfo {pages} {040903} (\bibinfo {year}
  {2020})}\BibitemShut {NoStop}%
\bibitem [{\citenamefont {Dukelsky}\ and\ \citenamefont
  {Pittel}(2004)}]{Dukelsky-2004}%
  \BibitemOpen
  \bibfield  {author} {\bibinfo {author} {\bibfnamefont {J.}~\bibnamefont
  {Dukelsky}}\ and\ \bibinfo {author} {\bibfnamefont {S.}~\bibnamefont
  {Pittel}},\ }\href {https://doi.org/10.1088/0034-4885/67/4/R02} {\bibfield
  {journal} {\bibinfo  {journal} {Reports on Progress in Physics}\ }\textbf
  {\bibinfo {volume} {67}},\ \bibinfo {pages} {513} (\bibinfo {year}
  {2004})}\BibitemShut {NoStop}%
\bibitem [{\citenamefont {Legeza}\ \emph {et~al.}(2015)\citenamefont {Legeza},
  \citenamefont {Veis}, \citenamefont {Poves},\ and\ \citenamefont
  {Dukelsky}}]{Legeza-2015}%
  \BibitemOpen
  \bibfield  {author} {\bibinfo {author} {\bibfnamefont {O.}~\bibnamefont
  {Legeza}}, \bibinfo {author} {\bibfnamefont {L.}~\bibnamefont {Veis}},
  \bibinfo {author} {\bibfnamefont {A.}~\bibnamefont {Poves}},\ and\ \bibinfo
  {author} {\bibfnamefont {J.}~\bibnamefont {Dukelsky}},\ }\href
  {https://doi.org/10.1103/physrevc.92.051303} {\bibfield  {journal} {\bibinfo
  {journal} {Physical Review C}\ }\textbf {\bibinfo {volume} {92}},\ \bibinfo
  {pages} {051303} (\bibinfo {year} {2015})}\BibitemShut {NoStop}%
\bibitem [{\citenamefont {Tichai}\ \emph {et~al.}(2023)\citenamefont {Tichai},
  \citenamefont {Knecht}, \citenamefont {Kruppa}, \citenamefont {Legeza},
  \citenamefont {Moca}, \citenamefont {Schwenk}, \citenamefont {Werner},\ and\
  \citenamefont {Zarand}}]{Tichai-2023}%
  \BibitemOpen
  \bibfield  {author} {\bibinfo {author} {\bibfnamefont {A.}~\bibnamefont
  {Tichai}}, \bibinfo {author} {\bibfnamefont {S.}~\bibnamefont {Knecht}},
  \bibinfo {author} {\bibfnamefont {A.}~\bibnamefont {Kruppa}}, \bibinfo
  {author} {\bibfnamefont {O.}~\bibnamefont {Legeza}}, \bibinfo {author}
  {\bibfnamefont {C.}~\bibnamefont {Moca}}, \bibinfo {author} {\bibfnamefont
  {A.}~\bibnamefont {Schwenk}}, \bibinfo {author} {\bibfnamefont
  {M.}~\bibnamefont {Werner}},\ and\ \bibinfo {author} {\bibfnamefont
  {G.}~\bibnamefont {Zarand}},\ }\href
  {https://doi.org/10.1016/j.physletb.2023.138139} {\bibfield  {journal}
  {\bibinfo  {journal} {Physics Letters B}\ }\textbf {\bibinfo {volume}
  {845}},\ \bibinfo {pages} {138139} (\bibinfo {year} {2023})}\BibitemShut
  {NoStop}%
\end{thebibliography}
%

\end{document}